\documentclass{aa}  
\usepackage{graphicx}
\usepackage{longtable,lscape}
\usepackage{lscape}
\usepackage{txfonts}
\begin{document}
\title{Emission line objects in NGC\,6822. New Planetary Nebula candidates.\thanks {Based on observations collected at Cerro Tololo Inter-American Observatory, National Optical Astronomy Observatory, which is operated by AURA, Inc., under cooperative agreement with the National Science Foundation.}
\thanks{Based on observations collected at the European Southern Observatory, VLT, Paranal, Chile, program ID 077.B-0430}}
\author{Liliana Hern\'andez-Mart\'inez\inst{1} \and Miriam Pe\~na\inst{1}
}
\offprints{L. Hern\'andez-Mart\'inez}
\institute{Instituto de Astronom{\'\i}a, Universidad Nacional Aut\'onoma de M\'exico, Apdo. P ostal 70264, M\'ex. D. F., 04510 M\'exico\\
\email{lhernand@astroscu.unam.mx, miriam@astroscu.unam.mx}
 }
\date{Received xxxxxxxxxxx; accepted xxxxxxxxxxxx}
\titlerunning{New PNe in NGC6822}
\authorrunning{Hern\'andez-Mart\'inez \& Pe\~na}


\abstract
{}
{Images obtained with the CTIO 4-m telescope and the MOSAIC-2 wide field camera in [O {\sc iii}] 5007 and H$\alpha$ on-band and off-band  filters are analyzed to search for emission lines objects in the dwarf galaxy NGC\,6822. In particular we search for Planetary Nebula (PN) candidates. In addition, 
imaging and spectroscopy of  a sub-sample of objects obtained with ESO VLT and FORS\,2 spectrograph are used to calibrate the MOSAIC imaging.}
{In  the continuum-subtracted  images, a large number of line emission regions were detected, for which we measured instrumental magnitudes in all the filters. The [O {\sc iii}] 5007 and H$\alpha$+[N {\sc ii}] magnitudes were calibrated with the spectroscopy.}
{Based upon some criteria to distinguish between PNe and compact HII regions, we have found 26 PN candidates, increasing the known sample in 8 objects. Also we detected a number of compact HII regions  and about 20 stellar-like objects emitting in  H$\alpha$. For all the objects  we present coordinates, instrumental magnitudes and nebular [O {\sc iii}] and H$\alpha$+[N {\sc ii}] fluxes.  The  observed luminosity function for the PN [O {\sc iii}] 5007 magnitudes (PNLF) and the cumulative PNLF were calculated. We confirm that the PNLF presents a dip similar to the one detected for the Small Magellanic Cloud at 2.5 mag down the maximum.  The cumulative PNLF returns a value $M^\star_{5007}=-3.71^{+0.21}_{-0.42}$ for the peak absolute magnitude of  the PNLF which is faint compared to the value expected for galaxies with metallicity similar to the one of NGC\,6822 but similar within uncertainties. From our best fit to the observed PNLF we obtained a rough estimate of the distance modulus
 m$-$M\,=\,23.64$^{+0.23}_{-0.43}$ mag, which agrees within uncertainties with recent values  from Cepheid stars reported in the literature.
Also the number of PN in the brightest 0.5 mag, normalized to the galactic bolometric luminosity, was estimated to be  $\alpha_{0.5} \sim  (3.8^{+0.90}_{-0.71})$ E-9. This number is similar to the values derived for galaxies with recent star formation and small galaxies (M$_{\rm B}$ fainter than $-$18 mag) and larger than the values for early-type galaxies.}
{}
\maketitle
\section{Introduction}

Planetary nebulae (PNe) are valuable tracers of the stellar population of low and intermediate mass stars. Due to their selective emission in a small number of strong and narrow emission lines, they can be discovered at significant distances within the nearby Universe (at least 30 Mpc). Their study provides accurate information on the luminosity, age, metallicity, and dynamics of the parent stellar population. This makes them very useful to test a number of theories about the evolution of stars and galaxies.

PNe are also useful as distance indicators through the  [O {\sc iii}] 5007 Planetary Nebulae Luminosity Function  (PNLF). The advantage of this method is that we can see bright PNe in galaxies of all kind of Hubble types, and PNe are easily indentified. Jacoby  (1989) and Ciardullo et al. (1987) reported that the PNLF had the same shape in all their investigated galaxies so it can be used as a standard candle, if we assume a complete sample in the 2 or 3 brightest magnitudes. However,  there is  evidence that the standard PNLF does not fit well in some galaxies. Jacoby \& De Marco (2002) show that the PNLF of the Small Magellanic Cloud presents  a prominent  dip  4 mag down from the brightest PN. Leisy et al. (2005), with 17 PN candidates, obtained  a  precarious PNLF for  NGC\,6822 which presents also a dip  around  2 mag down  the brightest PN. 
The main aim of this work is to investigate the PN populations in NGC 6822 by performing a deep survey for PN candidates.

NGC\,6822 (DDO 209, IC 4895) is a nearby Local Group gas-rich dIrr galaxy. It is located at  distance modulus of 23.31$\pm$0.02 (Cepheid distance by Gieren et al. 2006)  from our galaxy; moving away at V$_{\rm rad} \sim$ 44 km s$^{-1}$ and at 880 kpc from M31 (Mateo 1998).  Recently it has been shown that NGC\,6822 possesses two components: a huge  HI disk of about 6$\times$13 kpc, centered at the optical center, which includes the well-known optical bar and where the young stellar population resides extending in  zones with radii over  5 kpc from the center (de Blok \& Walter 2000; de Blok \& Walter 2006), and a second component constituted by a spheroidal stellar structure as extensive as the HI disk, but with its major axis at roughly right angles to it. This spheroid contains a substantial intermediate-age population (red giants, AGB  and carbon stars, Demers et al. 2006 and references therein).

 This galaxy has  clear evidence for recent star formation with more than a hundred HII regions detected (Killen \& Dufour 1982; Hodge et al. 1988). Ford et al. (2002) found 7 candidate PNe and more recently Leisy et al. (2005) have reported several  new  PN candidates, increasing the total sample  to 17, most of them located in the central zone. An additional PN was found by Richer \& McCall (2007). In the chemical context  NGC\,6822 is a metal-poor galaxy, with an interstellar medium (ISM) abundance  of about 0.2$ Z_\odot$ (e.g., Lee et al.  2006; Richer \& McCall 2007). Spectroscopic studies of some emission line objects has been performed by Killen \& Dufour (1982), Richer \& McCall (1995; 2007), Peimbert et al. 2005; and others. From these studies a number of PN candidates has been confirmed as true PNe.
 
Historically this galaxy seems to have not been  affected by tidal effects from the Milky Way or M31 thus it is suitable for chemical evolution studies. However, recent dynamical studies by de Blok \& Walter (2006),  reported a "North West companion", and other peculiar structures which indicate that probably there were at least some encounters or tidal effects in this galaxy.

     We obtained deep imaging of the whole face of NGC\,6822 in an effort of extending the known PN population towards  one or two magnitudes fainter. As part of the same program, a sub-sample of the detected PNe and HII regions  has been studied  spectroscopically to determine their chemical composition. First, we obtained on-band off-band images  in [O {\sc iii}] 5007 \AA~ and H$\alpha$, with wide field cameras, to search for emission line objects. We classified the objects according to the criteria  proposed by Pe\~na et al. (2007) and separated them in HII regions, PN candidates and H$\alpha$-emission stellar objects. In this paper we present the results of the imaging and use the spectra to validate the identification of the objects as PNe or HII regions. The results of spectroscopy will be presented in a second paper where line ratios, fluxes, physical conditions and   chemical compositions will be analyzed.
     
     The paper is organized as follows: in \S 2 we present the observations and data reduction. In \S 3 comparative photometry and flux calibration for the detected objects  are performed. The final samples and their distribution in the galaxy are presented in \S 4 and in \S 5 we discuss the differential PNLF, the cumulative PNLF, and the number of PNe normalized to the parent-galaxy bolometric luminosity, $\alpha_{0.5}$. Our results are summarized in \S 6.

  \section{Observations and data reduction}
  
\subsection{Imaging}
NGC\,6822 was observed on 2005 September 4 and October  9, using the CTIO 4-m  Blanco telescope with the MOSAIC\,2 camera. The NOAO CCD MOSAIC\,2 is a wide field imager of 8192 x 8192 pixels, with a spatial scale of 0.27$''$ per pix. This provides a field of view of 36.8$' \times 36.8'$. The whole image is divided in 8 CCDs, each one with two amplifiers.

Due to the angular size of the galaxy (about 28$' \times$ 40$'$ if we include  the distribution of faint C stars), it was possible to cover almost the entire galaxy  with just one exposure of the MOSAIC\,2 imager. This gave us the advantage of obtaining very homogeneous data in each filter. The center of the observed field is R.A. = 19:45:00.20,  Dec = $-$14:48:51.1. Filter characteristics and exposure times are listed in Table 1. In order to cover the CCD gaps between the 8 CCDs we took dithered exposures with the configuration (64, 126).

On 2006 August 20 and 22, we observed  NGC\,6822  with the ESO VLT UT1 (Antu) telescope, in Cerro Paranal, Chile. The FORS\,2 (Focal Reducer Spectrograph 2)  was used to obtain spectroscopic data to characterize the interstellar medium in NGC\,6822 (program ID 077.B-0480). The wide field of FORS\,2 (6.8$'  \times$ 6.8$'$) was used to study two  fields. The first one includes the central part of the galaxy centered in R.A.=19:44:55.5, Dec = $-$14:48:00.0 and the second  covers an outskirts zone in the north-west region of the galaxy centered in  R.A.=19:43:59.2, Dec = $-$14:46:14.9.  The  pre-imaging observations of the program, obtained to select the candidates for a subsequent multi-object spectroscopic run have been used here to corroborate the CTIO MOSAIC data. Filter characteristics and exposure times are presented in Table 1.

\begin{table}
\caption{CTIO-MOSAIC (above) and VLT-FORS\,2 (below) imaging characteristics}             
\label{table:1}      
\centering                          
\begin{tabular}{l c c c}        
\hline\hline                 
filter & $\lambda_c$ (\AA) & FWHM (\AA) & exp. time (s)  \\    
\hline                        
${\rm [O~III]}$ & 4990 & 50 & 8 $\times$ 1200 \\      
D51 & 5130 & 154 & 4 $\times$ 900 \\
H$\alpha$ & 6563 & 80 &  2 $\times$ 1200  \\
H$\alpha$+8 & 6650 & 80 & 2 $\times$ 1200 \\
\hline
${\rm [O~III]}$ & 5045 & 59 & 960 \\
V$_{\rm Bess}$ & 5540 & 1115 & 60 \\
H$\alpha$ & 6604 & 64 & 960 \\
R$_{\rm special}$ &  6550 & 1650 & 60 \\
\hline                                   
\end{tabular}
\end{table}

\subsection{Spectroscopy}
Spectrophotometric data for several PN candidates and compact HII regions were obtained with ESO Very LargeTelescope FORS\,2  in MXU mode, on 2006 August 20 and 21. The grisms 600B and 600RI were used to cover a spectral range from about 3600 to 7500 \AA.  The observed spectra are not affected by atmospheric dispersion because  the linear atmospheric dispersion compensator  (LADC) is automatically set  when observing with VLT-FORS. This compensates totally atmospheric dispersion effects when observing  
 at zenith distance lower than 45$^o$ which is our case. We obtained three frames for each grism, with exposure times of 1800\,s  for each 600B exposure and 1500\,s for each 600RI. The slit width was 1$''$ for all the objects and the spectral resolution varied from about 0.7 \AA ~ to 1.2 \AA.
Data were reduced and calibrated  with the ESO pipeline and IRAF\footnote{ IRAF is distributed by the National Optical Observatory, which is operated by the Associated Universities for Research in Astronomy, Inc., under contract to the National Science Foundation.} rutines. The standard stars EG274, LDS749B and BMP16274  were observed through a slit of 5$''$ width for flux calibration. In this paper we use the spectroscopic data to confirm the  classification of some objects and to calibrate the instrumental [O {\sc iii}] 5007 and H$\alpha$ magnitudes. During the spectroscopic run the sky was clear and the seeing conditions varied from 0.7 to 0.9$''$, however some flux could have been lost  in the 1$''$ slit. The losses can amount up to 15-20\% which should be considered in the uncertainties  of our flux calibration. The [O {\sc iii}] and H$\alpha$ spectroscopic fluxes are presented in Table 2.
In a second paper we will present all the spectral data: line fluxes, ratios, and their analysis.

\subsection{Data reduction}

The CTIO MOSAIC\,2 data were reduced using the MSCRED IRAF package. We followed the reduction procedure given by B. T. Jannuzi, J. Claver, and F.Valdes  in: http:// www.noao.edu/  noao/ noaodeep/ ReductionOpt/ framesv6. Html (Valdes 1998). We perform the cross talk correction, trimming, overscan, bias subtraction, flat field division, and the correction for bad pixels (bpm) and  cosmic rays (crmask) as indicated in this package. The astrometric solution was done using the USNO-A2.0 catalogue. Afterwards  we corrected by the tangent-plane projected, and substracted the sky.

The dithered [O {\sc iii}] 5007,  H$\alpha$  images and their off-band images were respectively shifted and combined in a single   8\,K$\times$8\,K image. Therefore, finally  we have a 2~hrs exposure time image in [O {\sc iii}] 5007, a 1~hr image  in off-band [O {\sc iii}], and 1~hr images for H$\alpha$ and its continuum. Observing nights were not photometric and the calibration flux could not be done with the standard star, but differential photometry can be performed. With the combination of seeing conditions and instrumental set-up we obtained a FWHM of 5.5 pix in H$\alpha$ image, 4.9 pix in H$\alpha$ continuum (hereafter Ha8); 3.9  pix in [O {\sc iii}] 5007, and 3.9 pix in D51. 

The ESO FORS\,2 pre-imaging were reduced and calibrated through the normal procedures of the ESO pipeline and the spectroscopy was reduced as mentioned in the previous section.

Our deep on-band  off-band CTIO and ESO images were subtracted to produce 2 difference images for detecting emission line objects. Also the ``blinking'' technique was used to identify objects. A number of extended and compact emitting  regions were found. All the objects detected in the [O {\sc iii}] images were also detected in the H$\alpha$ image, but additional stellar-like H$\alpha$ emitting objects were also found.
The complete sample has been marked on our CTIO H$\alpha$  image which is shown  in Fig.1. Different symbols represent  different type of objects (PNe, HII regions and  H$\alpha$ stellar objects). In the next section we discuss the criteria for distinguishing them. ID charts of the new objects are presented in the Appendix.
\begin{figure*}[ht] 
\begin{center}
\label{5007image}
\includegraphics[width=10cm,height=10cm]{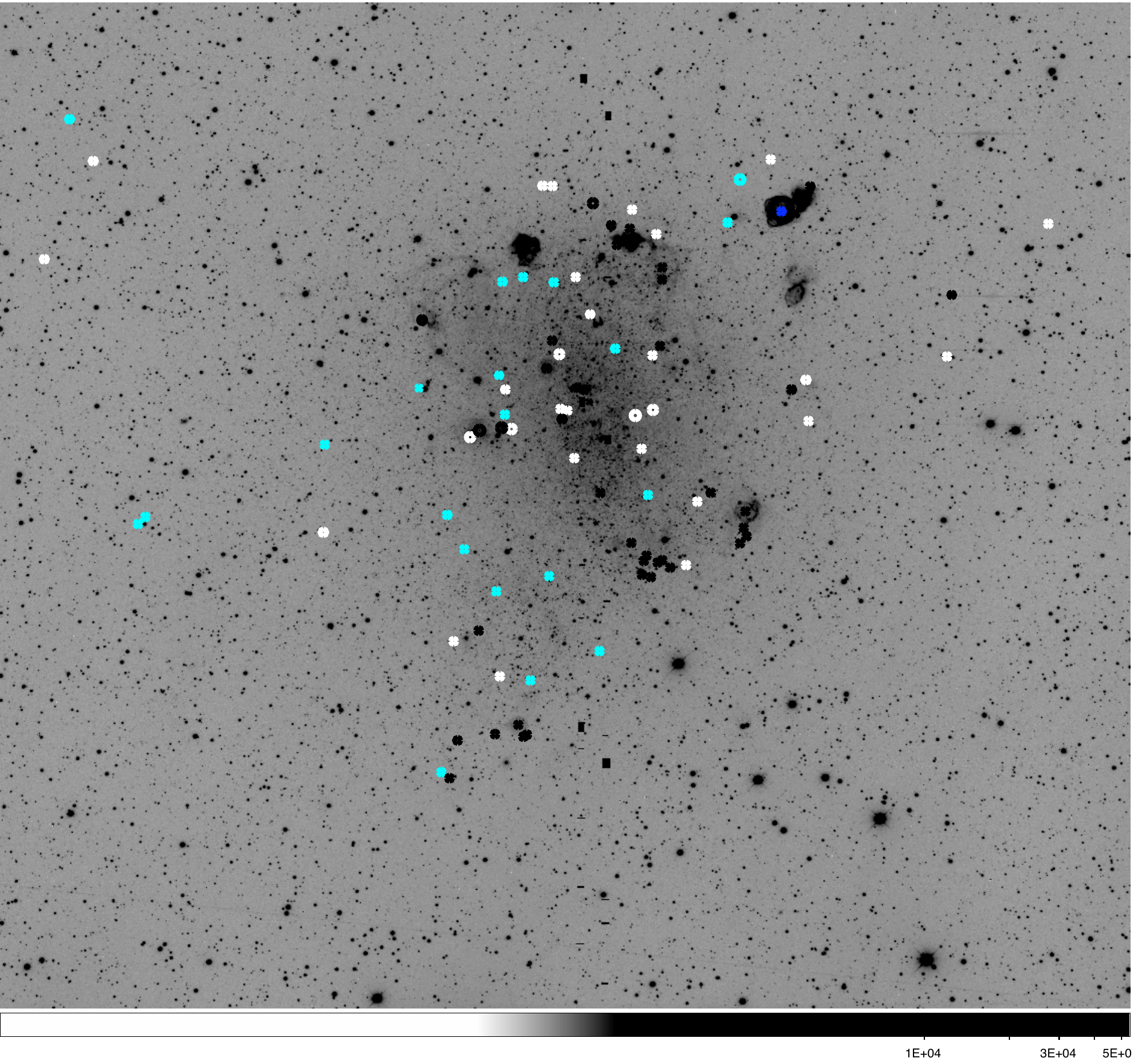}
\caption{Our CTIO MOSAIC\,2  H$\alpha$ image is shown. The field size is 36.8$\times$36.8 arcmin. North is up, East is left. The objects of Table 2 have been marked as follows: white circles represent PN candidates; black circles, HII regions; cyan circles, H$\alpha$ stellar objects. (See the text for a description of the selection mechanism). HII regions and H$\alpha$ stellar objects are concentrated in the central zones in very good coincidence with the HI disk by de Blok \& Walter (2000) while PNe are more widely distributed.}
\end{center}
\end{figure*}

\section{Distinguishing among  different types of emission line objects}

The way in which the PNe candidates were selected follows the prescriptions given in Pe\~na et al. (2007). That is, PN candidates should be point-like objects (at the distance of NGC\,6822, a 1 pc radius nebula  would appear as a 0.4$''$ size object, thus, stellar-like) and the central star should not be detected or it should be very faint as compared with the central stars in HII regions. This is considering that at the distance of NGC\,6822 an O9.5$-$B0 V star (the faintest able of producing a low excitation HII region), possessing an absolute magnitude $M_{\rm V}  \sim -3.90$, (Martins et al.  2005) has an apparent magnitude V$\sim$ 19.5 mag and a PN central star is typically 2$-$2.5 mag fainter (M\'endez et al. 1992). Certainly it could happen that a field star is projected on a nebula difficulting the classification. Only spectroscopy can help in such a situation (see the cases discussed by Richer \& McCall 2007). 

 The excitation degree, represented by the [O {\sc iii}] 5007\,/\,H$\alpha$ flux ratio, is a very used criterion to select PN candidates (e.g., Magrini et al. 2000; Ciardullo et al. 2002; Herrmann et al. 2008). It is based on the 
fact that  usually the brightest PNe have I(5007)\,/\,I(H$\alpha$) larger than HII regions.  Moreover, Ciardullo et 
al. (2002) showed that PNe inhabit a very distinctive region in the [O {\sc iii}]--H$\alpha$ space, with the brightest PNe showing dereddened  I(5007)\,/\,I(H$\alpha) \geq$ 2. Recently Herrmann et al. (2008) codified this criterion. In \S 3.2 we apply this criterion to our objects and discuss its validity for NGC\,6822.  

 In this section we will use the central star criterion to separate PN candidates from other type of emission line nebula.
To proceed we have performed  a comparative photometry measuring the instrumental magnitudes in the different MOSAIC  images:   [O {\sc iii}] 5007, D51, H$\alpha$ (Ha) and H$\alpha$+8 (Ha8). The IRAF task   digiphot.apphot.phot was used. An aperture of 7 pix radius (equivalent to 1.9$''$) was used to integrate the magnitudes of objects and  the sky was subtracted from a ring of 2$''$ width around. The results are listed in Table 2, where we present the coordinates (columns 2 and 3) and instrumental magnitudes (columns 4$-$7) for all the compact emission line objects. Note that due to the band width of H$\alpha$ filter, m(Ha) includes H$\alpha$ and both [N  {\sc ii}] 6548 \AA ~ and 6583 \AA ~ lines.  The instrumental photometric errors associated to the photometry are presented in Table 3.

\addtocounter{table}{1}
\begin{table}
\caption{Photometric errors associated to instrumental magnitudes}             
\label{table:3}      
\centering                          
\begin{tabular}{c c c c c}        
\hline\hline                 
mag & $\Delta$ m & & mag & $\Delta$ m  \\    
\hline                        
19.000 &	0.007 &&23.000 &	0.057 \\
19.500 &	0.008&&23.500 &	0.093\\
20.000 &	0.009&&24.000 &	0.117\\
20.500 &	0.012&&24.500 &	0.190\\
21.000 &	0.015&&24.800 &     0.215\\
21.500 &	0.021&&25.000 &	0.310\\
22.000 &	0.027&&25.500 &	0.530\\
22.500 &	0.035&&26.000 &	0.869\\
\hline                                   
\end{tabular}
\end{table}

The on-band magnitudes were plotted versus the off-band ones in Figures 2a and 2b. In these figures it is possible to select PN candidates (fill dots) as those nebulae possessing  a faint or undetectable star, while most HII regions (open squares) and H$\alpha$ stellar objects (small diamonds)  have  a much brighter m(D51) or m(H8). In Fig.~2a, m(5007) vs. m(D51), the emission nebulae (PNs and HIIs) show intense m(5007) magnitude while the H$\alpha$-emitting objects behave as normal stars. Some HII  regions are brighter than PN candidates and all of them have  brighter central stars. The two PNe, with a bright central star  at m(D51)$\sim$ 22 mag, are PN10 and PN20 in our list and they correspond to the previously known PN19 and PN20 for which Richer \& McCall (2007)  showed that a field star is projected on these nebulae. In Fig.~2b, m(Ha) vs. m(Ha8), the gap in stellar magnitudes  between HIIs and PNe is not as clear as in Fig.~2a  as there are three PNe  presenting a bright stellar continuum. They are  the objects PN4,  PN14 and PN20 of our list.
For PN20 we already mentioned that there is a star projected on the nebula; PN4 and PN14 are two very bright spectroscopically confirmed PNe (Hern\'andez-Mart{\'\i}nez \& Pe\~na, in preparation) where apparently a sort of red star is projected on the nebula.  
In general  HII regions are brighter in H$\alpha$ than PN candidates, as expected due to their brighter central stars.

In Fig. 2a)  the H$\alpha$ emitting stars follow closely the behavior of field stars, while in Fig. 2b) they also follows the line of stars  but at brighter m( H$\alpha$). The faintest among them could be low excitation PNe.

According with the results from Fig. 2, in Table 2 we have separated the objects by category: PN candidates, HII regions, stellar H$\alpha$ emission-line objects.

\begin{figure}[ht] 
\begin{center}
\label{magnitudes}
\includegraphics[width=9 cm,height=7cm]{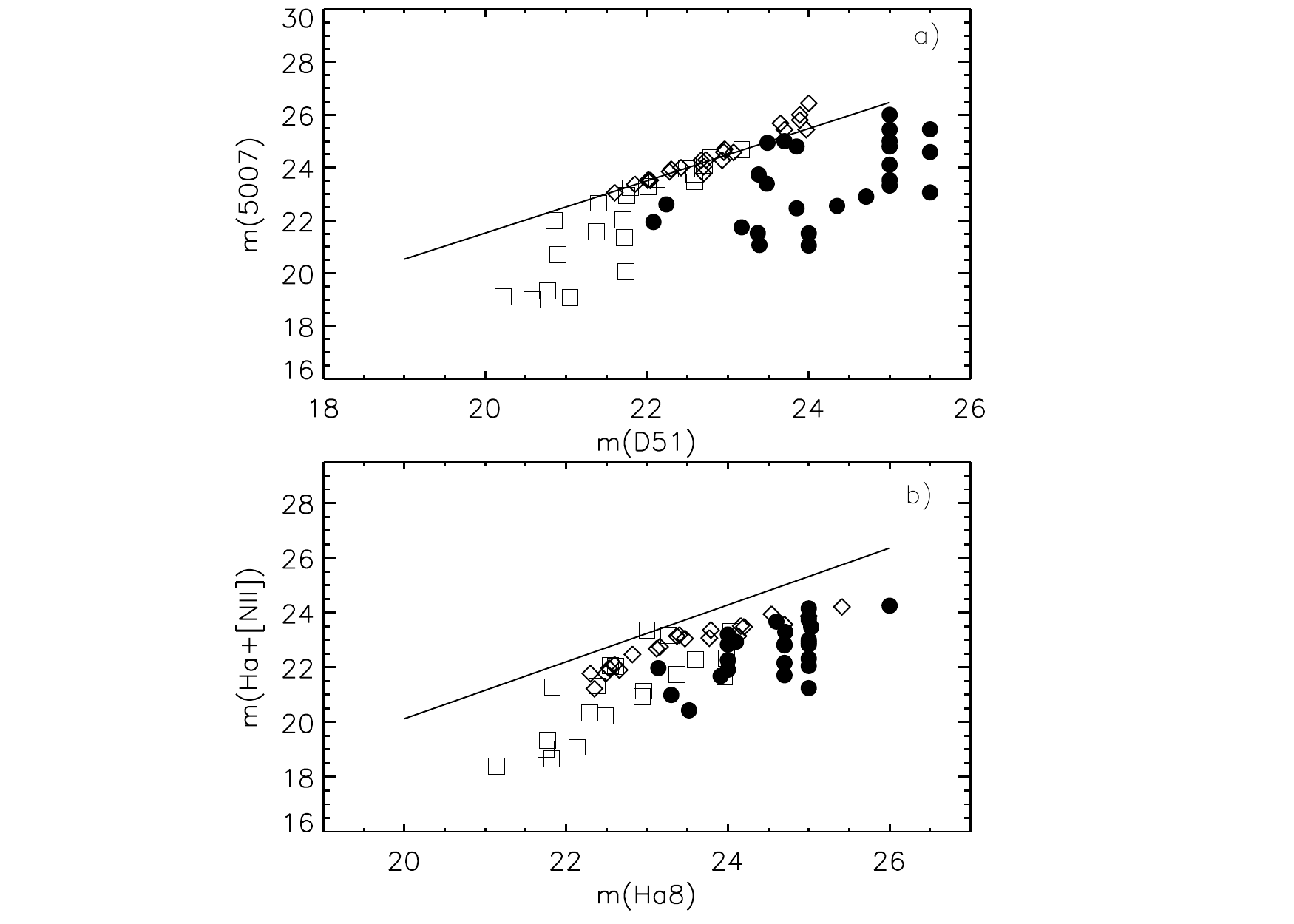}
\caption{On-band instrumental  magnitudes are plotted versus the off-band magnitudes. The solid line presents the behavior of field stars (with no line emission). PN candidates are marked with fill dots, HII regions with open squares and H$\alpha$  stellar objects as small diamonds. The accumulations of points at  m(D51) = 25 or 26 correspond to lower limits for m(D51). In Fig. a)  the H$\alpha$ emission stars follow closely the behavior of field stars, while in Fig. b) they also follows the line of stars  but at brighter m( H$\alpha$). The faintest among them could be low excitation PNe. Here the accumulation of points at m(Ha8) = 24, 25 and 26 is due to lower limits in m(Ha8).}
\end{center}
\end{figure}

\subsection{[O {\sc iii}] and H$\alpha$  Calibration} 

A more confident  way to distinguish between  different types of emission line objects  is by the spectral analysis of each object. However this is expensive in terms of telescope time because the objects are very faint. Our VLT spectroscopy included only 9 PNe and one  compact HII region. The calibrated [O {\sc iii}] 5007 \AA ~ and H$\alpha$+[N {\sc ii}] 6548,6583 \AA\AA ~  fluxes  measured for these objects are listed in columns 8 and 10 of Table 2. Figures 3a and 3b present the relation between our instrumental magnitudes and the logarithm of spectroscopic fluxes. It is notable  that a linear correlation can be fitted to both sets of  data with a very good correlation factor through 5 magnitudes. These fits (equations shown in the figure caption) are  used to calculate calibrated [O {\sc iii}] 5007 and H$\alpha$+[N {\sc ii}] fluxes for all the objects from their instrumental magnitudes (columns 9 and 11 of Table 2).   The error bars  in the fluxes (not shown in the table) include the errors in the photometry (Table 3) plus an uncertainty of about 20\% that should be added to the positive sign, due to possible losses in the spectroscopic slit (see \S 2). From the calibrated fluxes we have  calculated  the excitation degree [O {\sc iii}] 5007\,/\,H$\alpha$ for our objects (column 12 in Table 2), which is discussed in the next section.

\begin{figure}[ht] 
\begin{center}
\label{magnitudes}
\includegraphics[width=9cm,height=7cm]{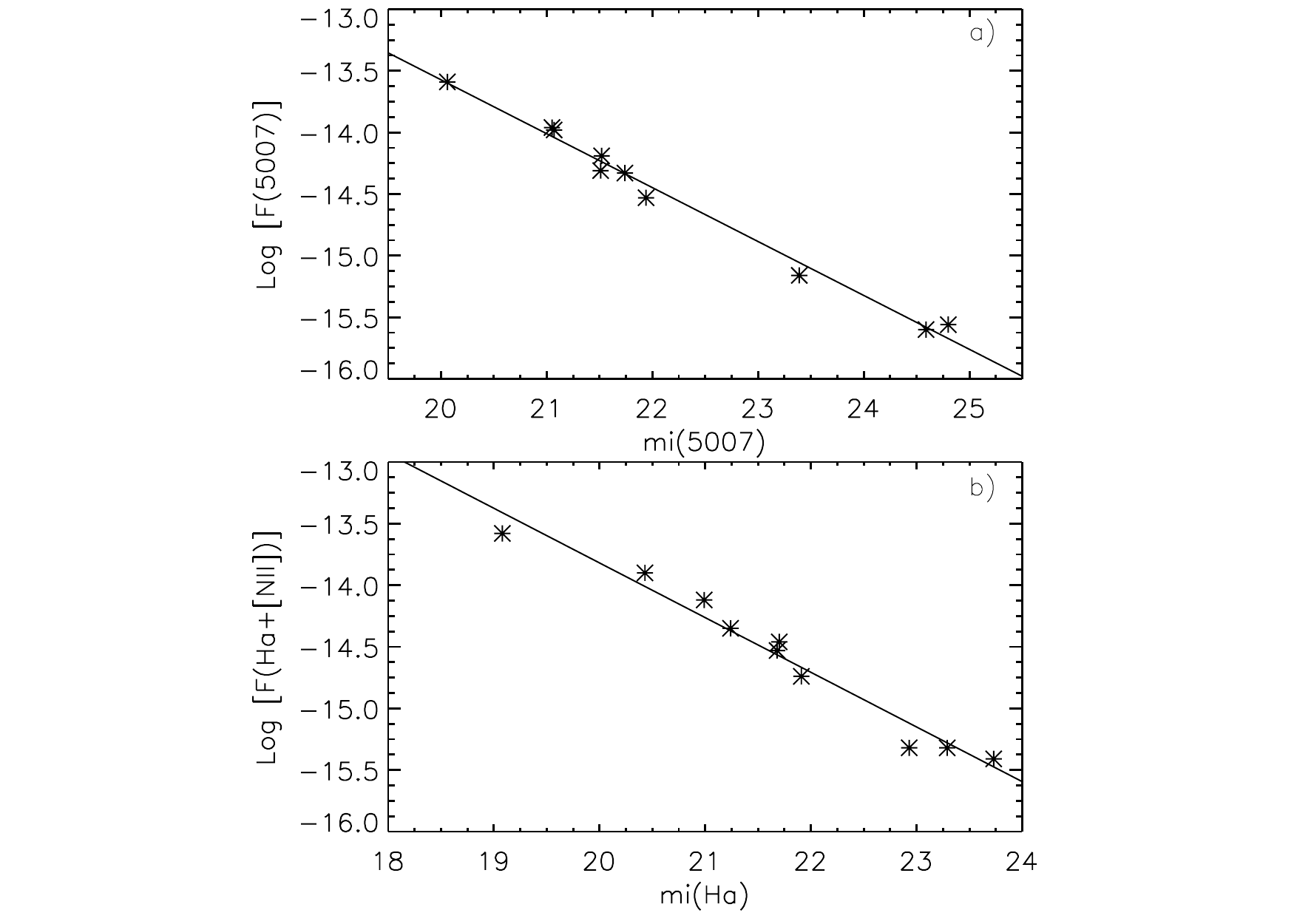}
\caption{The relations between the spectroscopic [O {\sc iii}] 5007 flux, F(5007), and the instrumental magnitude m$_{\rm i}$(5007)  (upper panel) and between the spectroscopic flux F(H$\alpha$+[N {\sc ii}]) and the instrumental  m(Ha) for the spectroscopically analyzed objects (lower panel). The linear fits correspond to: log F(5007) = $-$(4.82+0.44\,m$_{\rm i}$(5007)) in the upper panel (correlation coefficient R$^2$=0.95) and  log F(H$\alpha$+[N {\sc ii}]) = $-$(4.94\,+\,0.44\,m$_{\rm i}$(Ha)) in the lower panel  (correlation coefficient R$^2$=0.97). \label{fig3}}
\end{center}
\end{figure}

 \subsection{ The excitation degree of our emission line objects}

 In Fig. 4 we plot the log of the dereddened [O {\sc iii}]5007\,/\,(H$\alpha$+{N {\sc ii}]) flux ratio as a function of the absolute magnitude M(5007) for PN candidates and HII regions  of Table 2.  To deredden the apparent fluxes we have considered for NGC\,6822 a foreground extinction E(B-V)=0.26$\pm$0.04 mag (as derived for several authors, e.g., Massey et al. 1995; Gallart et al. 1996; Mateo 1998) and the reddening law by Fitzpatrick (1999). The Cepheids distance modulus,  23.31$\pm$0.02, was used to calculate the absolute magnitudes. The solid line represent the lower limit of the zone proposed by Ciardullo et al. (2002), and codified by Herrmann et al. (2008), as representative for  PNe in
the brightest 4 magnitudes, that is, the line corresponds to the equation:   ${\rm log }5007\,/\,H\alpha~>~ -0.37\, {\rm M}(5007) -1.16$.

\begin{figure}[ht] 
\begin{center}
\label{excitation}
\includegraphics[width=9.0 cm,height=7cm]{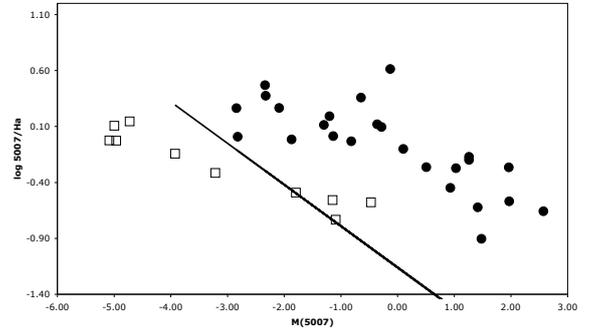}
\caption{The [O {\sc iii}]\,/\,H$\alpha$ flux ratio vs. the absolute magnitude M(5007) for PN candidates (black circles) and HII regions (open squares) in Table 2. The solid line corresponds to log 5007\,/\,H$\alpha = -0.37$\,M(5007) $-$ 1.16, which is the lower limit suggested by Herrmann et al. (2998) for the zone occupied by PNe in the brightest 4 magnitudes.}
\end{center}
\end{figure}

In this figure, it is evident that at a given I(5007)\,/\,I(Ha) the brightest objects are HII regions (as expected due to the larger number of ionizing photons of HII region central stars). At a given M(5007),  PNe always  have  larger 5007/H$\alpha$ ratios than HII regions and a ratio of 1.2 seems a safe value for most of the brightest PNe. All the PN candidates in our sample are safely above the limit by Herrmann et al., but  we should mention that  our brightest PN (PN10 in Table 2), which is a confirmed Type I PN with very intense [N II] (Richer \& McCall 2007; Hern\'andez-Mart{\'\i}nez 
\& Pe\~na in preparation)  have I(5007)\,/\,I(H$\alpha$+{N {\sc ii}]) lower than one due to the  [N {\sc ii}] lines. Therefore, an excitation-degree criterion requesting 5007-Ha $>$ 1.6, as the one suggested by Ciardullo et al. (2002), would  be discriminating against Type I PNe. 

In the faint  zone of Fig. 4,  where M(5007)$> -$2.0, all PN candidates  fulfill the criterion proposed by Herrmann et al. (2008) but  HII regions also fulfill this criterion. However HII regions can be easily separated from  PN candidates because they do not fulfill the other criteria: they do not have point-like appearance (at the distance of NGC\,6822 many HII regions  are diffuse) or they have a clearly visible central star. Then, for NGC\,6822, the limit proposed by Herrmann et al. helps to distinguish PNe from HII regions only in the 2 brightest magnitudes. For fainter objects definitely the faintness of the central star criterion is a better option.}}

\section{The  final samples and their distribution}

Finally,  from our deep search, we have a number of stellar and compact emitting objects which have been classified as PN candidates, HII regions and other stellar H$\alpha$ emitting objects (from the   characteristics of the latter ones, they could be  Ae-Be stars and a few could be low excitation PN) which are presented in Table 2. The list  includes coordinates, instrumental [O {\sc iii}] 5007, D51,  H$\alpha$+[N {\sc ii}] and Ha8 magnitudes, the [O {\sc iii}] and H$\alpha$+[N {\sc ii}]  fluxes for objects  spectroscopically observed, the fluxes derived from the calibration by using equations in Fig. 3a and 3b, and the excitation degree ([O {\sc iii}]/Ha). In the last column we present  previous IDs and comments.

The PN candidates amount 26, which is the largest sample obtained for NGC\,6822 so far. The PN confirmed by spectroscopy (13 objects) are marked with a {\bf  v} in the last column of Table 2. We rediscovered  all the 17 PN candidates reported by Leisy et al. (2005), the PN found by Richer \& McCall (2007) and all their HII regions.
 
 The distribution of emitting objects, differentiated by category, are shown in Fig. 1. Interestingly, all the samples are concentrated mainly in the central zones of the galaxy (coinciding with the optical bar and the main body of the HI disk reported by de Blok  \& Walter 2000), but HII regions and H$\alpha$ emitting objects are more concentrated than PN candidates. This indicates that the star-forming zones (characterized by HII regions) are very centrally concentrated with no recent star formation farther than  a few kpc  from the center. This result was also found by de Blok \& Walter (2006). On the other hand,  a few PN candidates are located  at much larger galactocentric distance. Possibly some of these PNe belong to the intermediate-age population of the spheroidal distribution  described by Demers et al. (2006). Anyway, there are large zones in the outer regions of the galaxy, in particular towards the South-West, where no PN has been found. This is mainly due to  the low stellar density  in this zone.

\section{The PN sample and the Planetary Nebulae Luminosity Function}

\begin{figure}[ht!] 
\begin{center}
\label{histo}
\includegraphics[width=9cm,height=7cm]{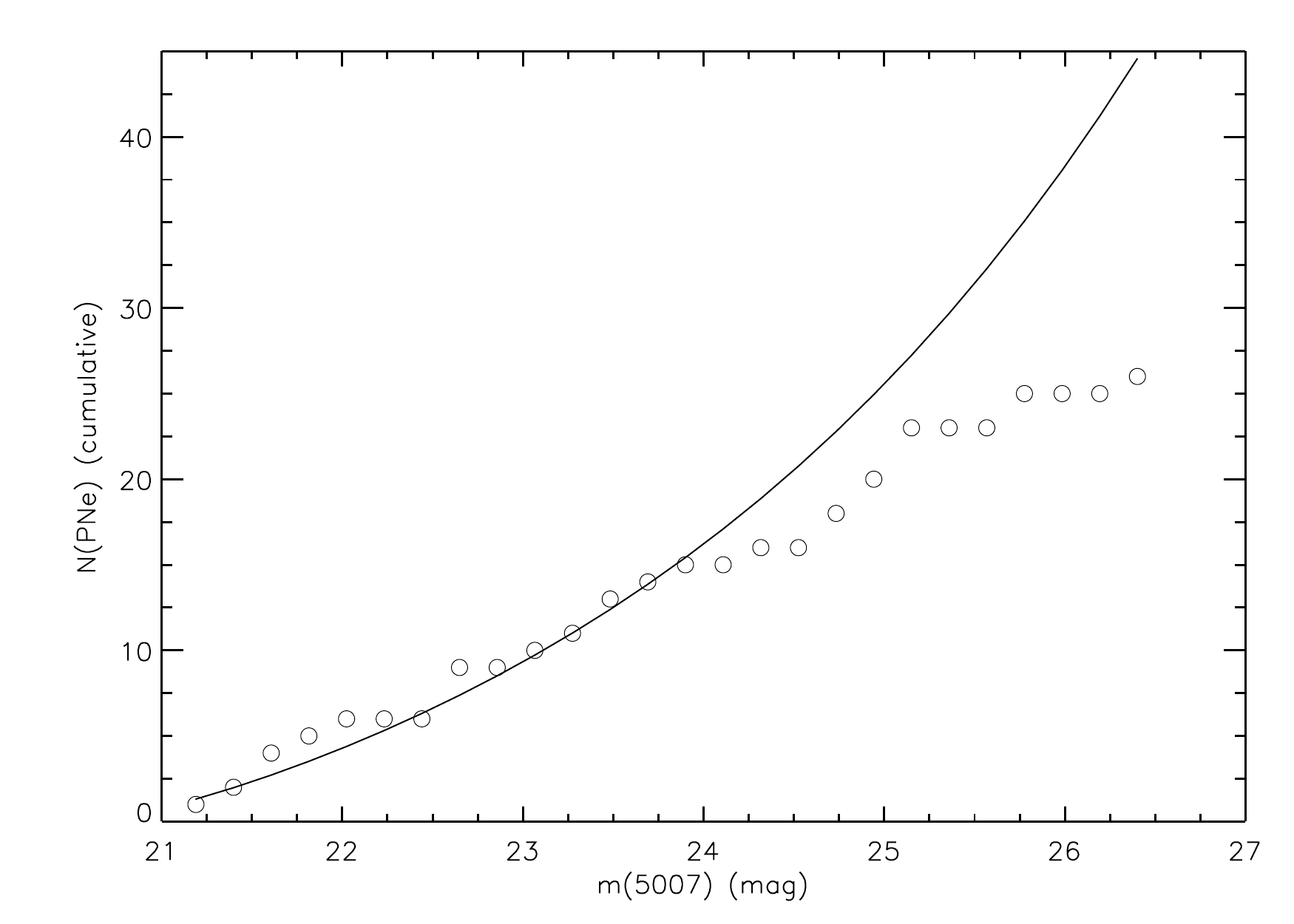}
\caption{Cumulative PNLF for NGC\,6822 PN candidates (data in Table\,2). The fit was  calculated for the 4 brightest magnitudes and it returns the product $N\exp(-0.307 \mu)$ and $m^*_{5007}$ (see text). }
\end{center}
\end{figure}

For our 26 PN candidates we determined [O {\sc iii}] 5007 and H$\alpha$+[N {\sc ii}] apparent fluxes. Then the [O {\sc iii}] 5007 luminosity function can be constructed. However, 26 objects is a very small number for a reliable statistics in order to derive the peak magnitude of the PNLF and  the distance to NGC\,6822. On the other hand, the PNLF helps for a better understanding of the PN populations  in a galaxy thus, with this purpose, in the following we will analyze first the cumulative PNLF and then the PNLF.

\subsection{The cumulative PN luminosity function}
The cumulative luminosity function can be constructed using apparent magnitudes instead of absolute magnitudes, as Pe\~na et al. (2007)  did for the PNe in NGC\,3109.  We convert  the usual luminosity function

N(M) $\propto \exp(0.307 M_{5007}) ~ (1-\exp(3(M^*_{5007}-M_{5007})))$    \hfill (1)
 
\noindent (Jacoby 1989; Ciardullo 1989) to a luminosity function in apparent magnitudes

N(m) = N $\exp(-0.307 \mu) ~\exp(0.307 m_{5007})~ (1-\exp(3(m^*_{5007} - m_{5007}))$\ , \hfill (2)

\noindent where N(m) is the number of objects brighter than m, N is a normalization constant, $\mu$ is the apparent distance modulus, $\mu = 5\, {\rm log\, d - 5 + A_{5007}}$, and $M^*_{5007}$ and $m^*_{5007}$ are the absolute and apparent peak magnitud of the luminosity function, respectively, the latter defined via 

m(5007) = $-$2.5 log F$_{5007} -$13.74 \hfill (3)

\noindent (Allen 1973; Jacoby 1989). 

 In Figure 5, we show the observed cumulative function  for PN data in Table 2. We can see a plateau at the faint end of the observed function, starting at about 25 mag. This is probably due to incompleteness of our sample. We have  fitted  eq. 2 to the first 4 mag of data using a fitting scheme based on the Levenberg-Marquardt technique which employes  a $\chi^2$ minimization.  The errors in the photometry,  as given in Table 3, were included in the fit procedure; therefore  the brightest magnitudes (with the smallest uncertainties) have larger weights in the fit.  The fit has been overlapped to the observed cumulative PNLF in Fig. 5. The best-fit parameters obtained are 
m$^*$(5007)= 20.47\,$\pm$\,0.17 and  N\,exp(-0.307$\mu)\,=\,(5.0\,\pm\,0.6) \times10^{-3}$. It should be mentioned  that the errors quoted for  these parameters (given at three  sigma level) correspond to the formal uncertainties of the fitting procedure and they do not represent real error bars because the latter ones are correlated.  A different statistical procedure  including a Monte-Carlo simulation is required to derive real errors (e.g., Hanes \& Whittaker 1987). Such a procedure is beyond the scope of this work.  

 For the foreground extinction at 5007 towards NGC\,6822 we adopted, as in \S 3.2, E(B-V)=0.26$\pm$0.04 mag which with Fitzpatrick (1999) reddening law corresponds to A$_{5007}$=\,3.34 E(B-V) = 0.87\,$\pm$\,0.13 mag. From this extinction and the distance modulus by Gieren et al.  (2006), we derive  $M^\star_{5007}=\,-3.71^{+0.21}_{-0.42}$.  These errors  were obtained from the formal errors given by the fit, the uncertainties in the extinction and the distance modulus by assuming that they add in quadrature. We have also included the  possibility that 20\% of the flux had been lost due to the narrow spectroscopic slit (thus objects could be 0.2 mag brighter). The value obtained for  $M^\star$(5007)   is faint compared with the value $-4.08$ predicted by Ciardullo et al. (2002) for galaxies with  a metallicity similar to the one of NGC\,6822, but the error bars are large in both cases and our result agrees with Ciardullo et al. one within uncertainties.

\subsection{The PNLF behavior}

 From our [O {\sc iii}] 5007 calibrated magnitudes for the PN sample we computed the observed  PNLF using a bin size of 0.78 mag which is presented in Figure 6.  In spite of the small number of objects, we are confident that the sample is complete in the 3 brightest magnitudes (a Kolmogorov-Smirnov test  was run to verify this, see below).
 
\begin{figure}[ht] 
\begin{center}
\label{PNLF}
\includegraphics[width=9cm,height=7cm]{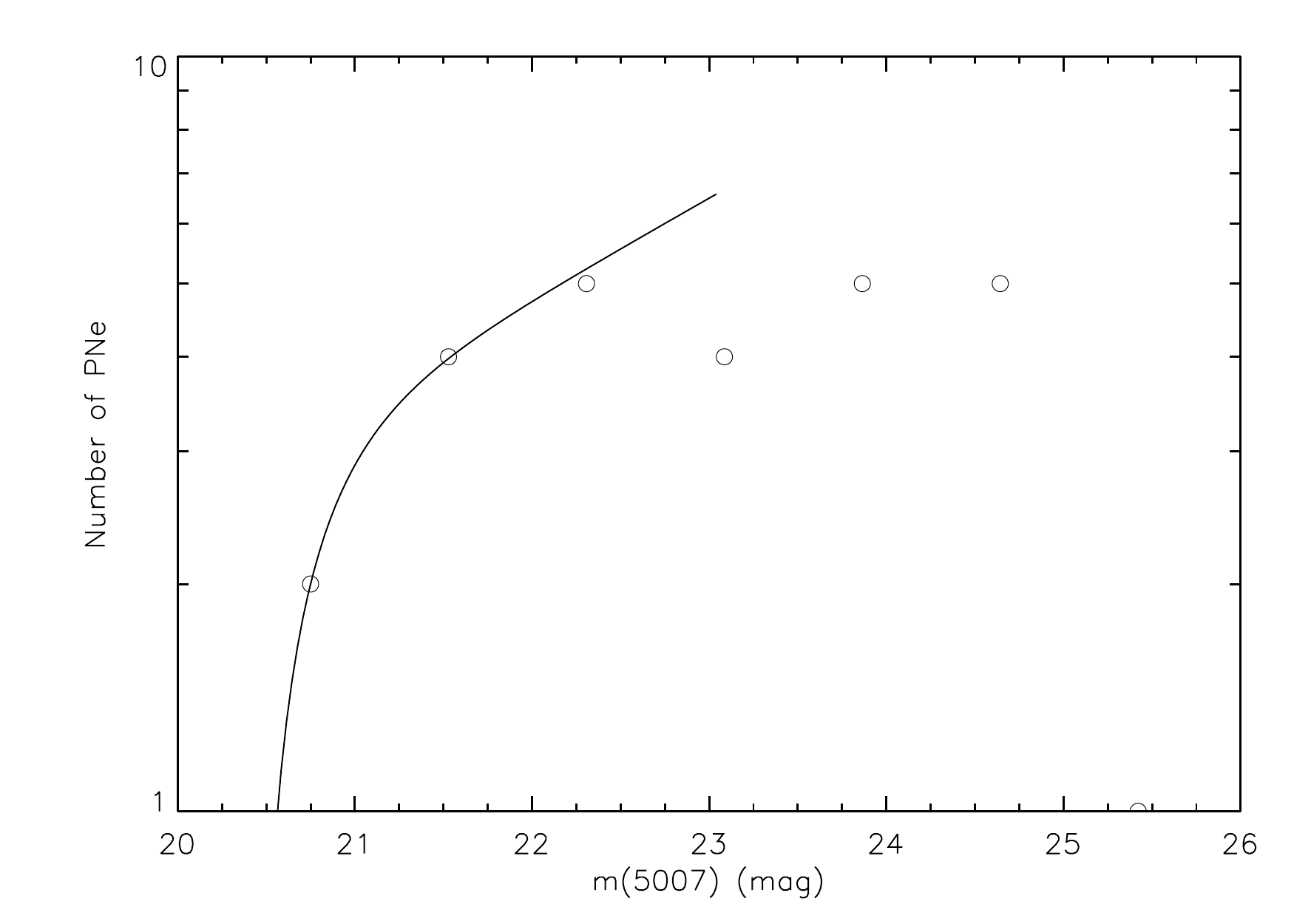}
\caption{The luminosity function for PNe in NGC\,6822.  A bin size of 0.78 mag has been used. The line is our best fit for the three brightest magnitudes. The PNLF shows a dip around 2.5 mag fainter than the brightest PN.}
\end{center}
\end{figure}

 Interestingly, the observed PNLF seems  not to increase monotonically with decreasing luminosity as predicted, but apparently it shows a dip at m(5007) $\sim$ 23 (about 2.5 magnitudes down the brightest PNe), similar to the one presented by the SMC PNLF (Jacoby \& De Marco 2005). This result was already mentioned by Leisy et al. (2005) with only 17 PN candidates. Our sample  includes 9  objects more; nevertheless, we did not find any new bright PNe apart from the ones  reported previously (Leisy et al. 2005; Richer \& McCall 2007). Thus, the dip seems statistically significant. To show this we  fit the empirical PNLF (eq. 1) 
 to the  3 brightest magnitudes (using the procedure given in the previous section and including  the photometric uncertainties of Table 3) and we run a Kolmogorov-Smirnov test in order to compare the observed PNLF with the empirical PNLF down to 23 magnitude (including the dip). The test shows that the observed PNLF  does not follow the empirical function at the 92\%    significance level, meaning that  the dip is statistically significant, however this result is not conclusive. 

 For the dip found in the SMC, Jacoby \& De Marco argued that this could be interpreted as due to the existence of a younger population of PNe in which the central star evolution proceeds very quickly. Such an interpretation could be valid for NGC\,6822 PNe, as this galaxy is similar in several  ways to the SMC and it seems to posses two stellar components.

As the top 3 mag of the PNLF are significantly complete and this is the distance-sensitive segment, it is tempting to make a rough estimate of the distance to NGC\,6822 via the fitting of this segment of the PNLF. From our best fit  we derive  an apparent peak 5007 magnitude m$^*$(5007) = 20.43$\pm$0.19 (error at three sigma). 
This value should be derredened for the foreground extinction. The same as before we used $A_{5007}=\,0.87\pm0.13$ mag. Then, by adopting a peak absolute magnitude M$^{\star}$(5007)= $-$4.08 (as derived from expression (3)  of Ciardullo et al. 2002 for NGC\,6822 metallicity), we obtain a distance modulus m$-$M=23.64$^{+0.23}_{-0.43}$. The error bars include the formal error of the fit, 
 the error in the reddening and, in the low side, a  20\%  due to possible flux losses in  the spectroscopic slit, has been added. Our derived distance   is  in  agreement, within uncertainties, with the distance  reported from Cepheids, e.g., 23.31$\pm$0.02 by Gieren et al. (2006).


\subsection{Computing $\alpha_{0.5}$}
An interesting parameter that can be estimated once the PNLF is known is $\alpha$, the number of PNe normalized to the parent-galaxy bolometric luminosity. Given the incompleteness of the PNLF at a few magnitudes below the maximum, Ciardullo et al. (2005) suggest to compute $\alpha_{0.5}$, that is, the ratio of PNe in the 0.5 brightest  magnitude to the bolometric luminosity.

From our observed PNLF, which we have tested it is complete in the first 2.5-3 magnitude, we find that the number of PNe in the first 2 mag is 6$_{-0.5}^{+1.0}$ objects. According to the integral PNLF   there should be 9 times more PNe in the top 2 mag than in the top 0.5 mag. Therefore the number of PNe in the brightest 0.5 mag is 0.67$_{-0.06}^{+0.11}$. 
The bolometric luminosity can be estimated from the absolute visual magnitude and by assuming a bolometric correction according with the galactic type (Buzzoni et al. 2006). From Mateo (1998) we adopted the visual integrated apparent magnitude to be V= 9.1$\pm$0.2 mag, which can be dereddened with E(B-V)\,=\,0.26\,$\pm$\,0.04 mag (A$_{\rm V}$ = 3.04 x E(B-V)= 0.79), giving V$_0$=\,8.3\,$\pm$\,0.2. Then the  absolute M$_{\rm V}$ is $-15.0\pm0.2$ if the Cepheid distance modulus 23.31$\pm0.02$  is used. Buzzoni et al. (2006)
suggest a mean bolometric correction of $-$0.8 as representative for any kind of galaxy; in particular for irregular galaxies this value could be slightly smaller, $-0.9$ (see their Fig. 9). We will adopt $-0.9\pm0.1$. Thus the bolometric magnitude for NGC\,6822 is about  $-15.9\pm0.22$ mag, resulting in a bolometric luminosity L$_{\rm bol}$=1.77E8$_{-0.34}^{+0.41}$ L$_\odot$.

 Thus we obtained $\alpha_{0.5} \sim  (3.8^{+0.90}_{-0.71})$ E-9. It has been found that $\alpha_{0.5}$  depends upon some properties of the parent galaxy. Galaxies with recent star formation and small galaxies show larger $\alpha_{0.5}$ than early-type galaxies (Ciardullo et al. 2005). The value derived for NGC\,6822  is similar to  the values derived for small galaxies (M$_{\rm B}$ fainter than $-$18) and galaxies with recent star formation, which have  $\alpha_{0.5} \sim 2.5{\rm E}-9$. In this sense the PN density to bolometric luminosity in NGC\,6822 behaves accordingly to its classification as a dwarf irregular.


\section{ Conclusions}
Deep [O {\sc iii}] 5007 and H$\alpha$+[N {\sc ii}] on-band off-band imaging have been performed with wide field cameras (in particular CTIO MOSAIC\,2) to search  line emission  objects in the dwarf irregular galaxy NGC\,6822.
For some objects we obtained follow-up spectroscopy during a spectroscopic run with  VLT FORS\,2. Our results are the following:

1) Using several criteria to distinguish between different types of line emission  objects we detected 26 PN candidates (8 more than the previously known sample), many compact HII regions and some H$\alpha$ emitting stellar objects (the brightest of them could be Ae-Be stars). From our follow-up spectroscopy  and using results from the literature, 13 of the 26 PN candidates have been confirmed as true planetary nebulae.

2) We find that compact HII regions and H$\alpha$ emitting objects are distributed in the central zone of the galaxy where also resides the  HI disk found by de Blok \& Walter (2000) and the young stellar population. No recent star-forming region is found  farther than a few kpc from the center. On the other hand, PNe  are more widely distributed, with some objects lying in the extreme west and east of the galaxy. This indicates that PNe belong to a different more widely spread stellar population. 

3) The instrumental [O {\sc iii}] 5007 and H$\alpha$+[N {\sc ii}] magnitudes derived from the imaging were calibrated using results from our spectroscopy. This allowed us to determine calibrated apparent  [O {\sc iii}] 5007 and H$\alpha$+[N {\sc ii}] magnitudes.

 4) We built the observed cumulative and  differential  [O {\sc iii}] PNLF. The differential PNLF presents a dip at about 2.5 mag down the brightest magnitude, similar to the one found in the SMC by Jacoby \& De Marco (2002). A Kolmogorov-Smirnoff test comparing the  observed and empirical PNLFs  shows that this dip is statistically significant with a 92\% significance level.  As in the case of the SMC, the dip can be explained  arguing that it could be caused by the presence of two different  PN populations where the younger would have central stars evolving very quickly. As NGC\,6822 seems to be a  galaxy with two stellar components: a HI disk with young population and a spheroidal stellar component, two PN populations could be expected in this galaxy. 
 
5) Our best fit for the 3 brightest magnitudes of the observed PNLF, allows us to determine a distance modulus  
m$-$M=23.64$^{+0.23}_{-0.43}$ mag which, within uncertainties, is in  agreement with recent values  from Cepheid stars reported in the literature.

6)  From the cumulative PNLF we derived the peak absolute magnitude of  the PNLF, $M^\star_{5007}=-3.71^{+0.21}_{-0.42}$ which is faint compared with the value $-4.08$ predicted by Ciardullo et al. (2002) for galaxies with  a metallicity similar to the one of NGC\,6822, but the error bars are large  and our result agrees with Ciardullo et al. one within uncertainties.
One of the benefit of observing low-metallicity galaxies like NGC\,6822 is a better understanding of the absolute magnitudes of their planetary nebulae.

 7) We have estimated the number of PNe in the brightest 0.5 mag, normalized to the galactic bolometric luminosity, $\alpha_{0.5}$, to be  $(3.8^{+0.90}_{-0.71})$ E-9. This value is similar to the values derived for small galaxies (M$_{\rm B}$ fainter than $-$18 mag) and galaxies with recent star formation, and larger than the values obtained for early-type galaxies. Thus the PN density to bolometric luminosity in NGC\,6822 behaves accordingly to its classification as a dwarf irregular.

\begin{acknowledgements}
Invaluable comments and support by  Michael Richer are deeply appreciated.  M. Pe\~na is grateful to DAS, Universidad de Chile, for hospitality during a sabbatical stay when part of this work was performed.  L. H.-M. benefited from the hospitality of the Departamento de Astronom{\'\i}a, Universidad de Chile. L. H.-M. received scholarship from  CONACYT-M\'exico and DGAPA-UNAM. 
 M. P. gratefully acknowledges financial support from FONDAP-Chile and DGAPA-UNAM. This work received financial support from grants \#43121 (CONACYT-M\'exico),  IN-114805 and IN-112708 (DGAPA-UNAM).
  \end{acknowledgements}

%
\longtabL{2}{
\begin{landscape}
\begin{longtable}{cccrcccccrrll}
\caption{Characteristics of  emission line objects in NGC\,6822 \label{tbl-1}}\\
\hline\hline
obj No.  &  R. A.$^2$   &  Dec$^2$  &  m$_i$(5007)  &  m$_i$(D51) &  m$_i$(Ha) &  m$_i$(Ha8)  &  log F$_{5007}^3$  & log F$_{5007}^4$  & log F$_{Ha}^3$  & Log F$_{Ha}^5$ & 5007/Ha  & other ID$^6$, comments  \\
\hline
\endfirsthead
\caption{continued.}\\
\hline\hline
obj No.  &  R. A.$^2$  &  Dec$^2$  &  m$_i$(5007)  &  m$_i$(D51) &  m$_i$(Ha) &  m$_i$(Ha8)  &  log F$_{5007}^2$  & log F$_{5007}^3$  & log F$_{Ha}^2$  & Log F$_{Ha}^4$ & 5007/Ha  & other ID$^5$, comments  \\\hline
\endhead
\hline
\endfoot
PN 1 & 19:44:35.45\footnote{Coordinates for equinox 2000.} & -14:40:50.5 & 24.94 & 23.49 & 22.84 & 24.70 & \footnote{Flux from spectroscopy, in erg cm$^{-2}$ s$^{-1}$. Ha= H$\alpha$+[N{\sc ii}]} & -15.73\footnote{Flux calculated from equation in Fig. 3a, in erg cm$^{-2}$ s$^{-1}$.} &  & -14.99\footnote{Flux calculated from equation in Fig. 3b, in erg cm$^{-2}$ s$^{-1}$.} & 0.18 & faint  \\
PN 2 & 19:45:56.38 & -14:40:50.5 & 22.55 & 24.35 & 22.32 & 25.00 &  & -14.69 &  & -14.76 & 1.18 & PN5\footnote{Previous ID from Leisy et al. 2005, Richer \& McCall 2007 and Killen \& Dufour 1982.}\\
PN 3 & 19:45:02.70 & -14:41:36.0 & 24.50 & --- & 22.80 & 24.70 &  & -15.54 &  & -14.97 & 0.27 & faint 5007\\
PN 4 & 19:45:01.53 & -14:41:36.3 & 21.05 &  $>$ 24.0 & 20.99 & 23.30 & -13.96 & -14.03 & -14.12 & -14.17 & 1.39 & PN4,  v\\
PN 5 & 19:44:52.02 & -14:42:18.0 & 24.11 & $>$ 25.0 & 22.83 & $>$ 25.0 &  & -15.37 &  & -14.99 & 0.41 & PN3,  v\\
PN 6 & 19:44:02.29 & -14:42:43.3 & 21.52 & 23.37 & 21.71 & 24.70 & -14.19 & -14.24 & -14.46 & -14.49 & 1.80 & PN1,  v\\
PN 7 & 19:44:49.11 & -14:43:00.6 & 22.46 & 23.85 & 22.05 & $>$  25.0 &  & -14.65 &  & -14.64 & 0.99 & PN2,  v\\
PN 8 & 19:46:02.22 & -14:43:42.2 & 22.90 & 24.71 & 22.16 & 24.70 &  & -14.84 &  & -14.69 & 0.70 & PN6\\
PN 9 & 19:44:58.77 & -14:44:14.7 & $>$ 25.0 & 23.70 & 22.26 & $>$ 24.0 &  & -15.76 &  & -14.74 & 0.09 & faint faint CS \\
PN 10 & 19:44:56.93 & -14:45:18.6 & 21.94 & 22.08 & 21.24 &   $>$  25.0 & -14.53 & -14.42 & -14.35 & -14.29 & 0.73 & PN19,  v \\
PN 11 & 19:45:00.74 & -14:46:28.9 & 25.45 &   $>$  25.5 & 23.47 & 25.03 &  & -15.96 &  & -15.27 & 0.20 & faint \\
PN 12 & 19:44:49.59 & -14:46:31.3 & 21.51 & $>$ 24.0 & 21.91 & $>$ 24.0 & -14.31 & -14.23 & -14.74 & -14.58 & 2.23 & PN14,  v \\
PN 13 & 19:44:31.27 & -14:47:14.9 & 23.74 & 23.38 & 22.83 & $>$ 24.0 &  & -15.21 &  & -14.99 & 0.60 &  \\
PN 14 & 19:45:07.15 & -14:47:31.4 & 21.07 & 23.39 & 20.43 & 23.52 & -13.98 & -14.04 & -13.90 & -13.93 & 0.77 & PN7,  v \\
PN 15 & 19:45:00.59 & -14:48:04.6 &   $>$ 25.3 &   $>$  25.0 & 23.40 & 25.24 &  & -15.87 &  & -15.24 & --- & no 5007 \\
PN 16 & 19:44:49.52 & -14:48:03.5 & 21.74 & 23.17 & 21.68 & 23.91 & -14.33 & -14.33 & -14.53 & -14.48 & 1.40 & PN13,  v  \\
PN 17 & 19:44:59.73 & -14:48:07.2 & 23.06 &   $>$ 25.5 & 23.20 & $>$ 24.0 &  & -14.91 &  & -15.15 & 1.73 & PN10 \\
PN 18 & 19:44:51.792 & -14:48:15.7 & 24.59 &   $>$ 25.5 & 23.29 & 24.71 & -15.60 & -15.58 &  & -15.19 & 0.41 & PN11,  v \\
PN 19 & 19:45:06.44 & -14:48:39.5 & 24.80 & $>$ 25.0     & 23.73 & $>$ 25.0 & -15.56 & -15.67 & -15.41 & -15.38 & 0.51 & PN9,  v \\
PN 20 & 19:45:11.5 & -14:48:53.6 & 22.61 & 22.24 & 21.97 & 23.14 &  & -14.71 &  & -14.61 & 0.78 & PN20,  v \\
PN 21 & 19:44:50.9 & -14:49:13.8 & 23.39 & 23.48 & 22.93 & 24.10 & -15.16 & -15.06 & -15.32 & -15.03 & 0.94 & PN12,  v \\
PN 22 & 19:44:58.9 & -14:49:31.0 & 24.80 & 23.85 & 23.67 & 24.60 &  & -15.67 &  & -15.35 & 0.48 & faint \\
PN 23 & 19:44:44.22 & -14:50:46.4 & 23.32 & $>$ 25.0 & 22.92 & $>$ 25.0 &  & -15.02 &  & -15.02 & 1.00 & PN17 \\
PN 24 & 19:44:45.59 & -14:52:37.0 & 23.53 &$>$  25.0 & 24.25 & $>$ 26.0 &  & -15.12 &  & -15.61 & 3.12 & PN16 \\
PN 25 & 19:45:13.36 & -14:54:48.6 & 25.44 & $>$ 25.0 & 24.15 & $>$ 25.0 &  & -15.95 &  & -15.57 & 0.41 & faint 5007 \\
PN 26 & 19:45:07.84 & -14:55:50.3 & $>$ 26.0 & $>$ 25.0 & 23.81 & $>$ 25.0 &  & -16.20 &  & -15.42 & --- & no 5007 \\
\hline 
HII 1 & 19:44:30.74 & -14:41:37.8 & 19.11 & 20.22 & 18.40 & 21.14 &  & -13.18 &  & -13.04 & 0.71 & \\
HII 2 & 19:44:56.8 & -14:42:07.9 &   $>$ 25.0 & 22.15 & 22.48 & --- &  &   $<$  -15.74 &  & -14.83 &   --- & HII\,21,  dif,   no CS\\
HII 3 & 19:44:32.54 & -14:42:23.8 &$>$ 24.5 & 22.58 & 21.74 & 23.37 &  & $<$-15.32 &  & -14.50 & --- & comp,  no 5007\\
HII 4 & 19:44.54.52 & -14:42:44.4 & $>$24.5 & 22.01 & 20.92 & 22.94 &  & $<$-15.32 &  & -14.15 & --- & HI\,20, comp,  no 5007\\
HII 5 & 19:44:52.31 & -14:42:50.6 & $>$24.5 & 22.79 & 22.27 & 23.60 &  & $<$-15.48 &  & -14.74 & --- & HII\,22,  dif,  no5007\\
HII 6 & 19:44:53.99 & -14:43:12.5 & 23.71 & 20.85 & 21.28 & 21.83 &  & -14.98 &  & -14.30 & 0.20 & dif,   faint \\
HII 7 & 19:44:48.41 & -14:43:58.6 & 21.57 & 21.37 & 22.03 & 22.61 &  & -14.26 &  & -14.63 & --- & dif, bright star \\
HII 8 & 19:44:58.50 & -14:44:45.7 & $>$24.5 & 23.07 & 21.76 & $>$ 25.0 &  & -15.40 &  & -14.51 & $<$ 0.13 & HII\,24, dif,  faint 5007\\
HII 9 & 19:44:13.81 & -14:44:46.9 & $>$24.5 & 22.59 & 21.66 & 23.96 &  & $<$-15.32 &  & -14.47 & ---  &  dif,  no 5007 \\
HII 10  & 19:45:17.09 & -14:45:29.5 & 22.01 & 21.70 & 20.23 & 22.48 &  & -14.45 &  & -13.84 & 0.24 & dif,   faint 5007\\
HII 11 & 19:45:01.58 & -14:46:05.6 & $>$25.0 &   $>$  25.0 & 23.07 & 24.07 &  & $<$-16.00 &  & -15.09 & --- & dif,  only Ha \\
HII 12  & 19:44:48.69 & -14:46:15.0 & 23.24 & 21.79 & 23.35 & 23.00 &  & -14.99 &  & -15.21 & --- & dif, bright star \\
HII 13 & 19:44:32.99 & -14:47:31.6 & 19.08 & 21.05 & 18.67 & 21.82 &  & -13.17 &  & -13.16 & 0.97 & KD\_H$\beta$, comp \\
HII 14 & 19:44:57.15 & -14:47:50.0 & 20.06 & 21.74 & 19.08 & 22.14 & -13.60 & -13.57 & -13.58 & -13.34 & 0.55 & HII\,08 \\
HII 15 & 19:44:30.88 & -14:48:26.3 & 19.00 & 20.58 & 18.30 & --- &  & -13.13 &  & -12.99 & 0.72 & KD\_S2,  comp \\ 
HII 16 & 19:44:55.79 & -14:50:31.3 & $>$24.5 & 21.75 & 22.07 & 22.55 &  & $<$ -15.32 &  & -14.65 & --- & dif,   no 5007\\
HII 17 & 19:44:42.57 & -14:50:32.2 & 21.35 & 21.72 & 20.34 & 22.29 &  & -14.16 &  & -13.89 & 0.53 & comp \\
HII 18 & 19:44:50.6 & -14:52:29.8 & $>$24.5 & 22.12 & 23.17 & 23.27 &  & -15.13 &  & -15.13 & --- & dif  \\
HII 19 & 19:44:50.8 & -14:52.53.5 & 19.33 & 20.77 & 19.01 & 21.75 &  & -13.28 &  & -13.30 & 1.06 &   comp \\
HII 20 & 19:45:10.37 & -14:54:30.4 & $>$ 24.5 & 22.49 & 21.13 & 22.96 &  & -15.30 &  & -14.24 & --- & comp, faint 5007 \\
HII 21 & 19:45:08.44 & -14:57:30.6 & $>$ 24.5 & 23.17 & 22.34 & 23.98 &  & -15.62 &  & -14.77 & --- & dif,  faint\\
HII 22 & 19:45:04.60 & -14:57:31.7 & 20.71 & 20.90 & 19.34 & 21.77 &  & -13.88 &  & -13.45 & 0.37 & comp  \\
HII 23 & 19:45:12.9 & -14:57:41.4 & $>$ 25.0 & 22.71 & 23.30 & 24.04 &  &  $<$ -15.8   &  & -15.19 & --- & dif, faint CS\\
HII 24 & 19:45:13.9 & -14:58:47.2 & 22.65 & 21.40 & 21.32 & 22.38 &  & -14.73 &  & -14.32 & 0.39 & dif \\
\hline  
Ha 1 & 19:45:59.21 & -14:39:37.6 & 25.80 & 23.89 & 23.56 & 24.70 &  & -16.11 &  & -15.30 & 0.16 &  faint PN?\\
Ha 2 & 19:44:39.1 & -14:41:25.7 & 25.68 & 23.65 & 23.87 & 23.74 &  & -16.06 &  & -15.44 & 0.24\\
Ha 3 & 19:44:40.62 & -14:42:40.7 & 23.94 & 22.30 & 22.75 & 23.16 &  & -15.30 &  & -14.95 & 0.45\\
Ha 4 & 19:45:05.02 & -14:44:14.8 & 24.59 & 22.95 & 23.14 & 24.13 &  & -15.58 &  & -15.12 & 0.35\\
Ha 5 & 19:45:07.51 & -14:44:22.2 & 24.04 & 22.70 & 21.22 & 23.35 &  & -15.34 &  & -14.28 & 0.09\\
Ha 6 & 19:45:01.36 & -14:44:23.9 & 23.78 & 22.70 & 21.97 & 22.53 &  & -15.23 &  & -14.61 & 0.24\\
Ha 7 & 19:44:54.04 & -14:46:20.0 & 23.37 & 21.85 & 22.47 & 22.82 &  & -15.05 &  & -14.83 & 0.60\\
Ha 8  & 19:45:07.95 & -14:47:05.6 & 24.29 & 22.73 & 21.90 & 22.66 &  & -15.45 &  & -14.58 & 0.13\\
Ha 9 & 19:45:17.50 & -14:47:27.6 & 23.52 & 22.01 & 21.77 & 22.49 &  & -15.11&  & -14.52 & 0.26\\
Ha 10 & 19:45:07.24 & -14:48:14.3 & 24.70 & 22.96 & 23.07 & 23.77 &  & -15.63 &  & -15.09 & 0.29\\
Ha 11 & 19:45:28.73 & -14:49:05.6 & 24.28 & 22.93 & 23.48 & 24.20 &  & -15.44 &  & -15.27 & 0.67 & faint PN?\\
Ha 12 & 19:44:50.12 & -14:50:35.2 & 24.00 & 22.42 & 23.05 & 23.47 &  & -15.32 &  & -15.08 & 0.57\\
Ha 13 & 19:45:14.11 & -14:51:09.9 & 23.05 & 21.60 & 22.09 & 22.60 &  & -14.91 &  & -14.66 & 0.57\\
Ha 14 & 19:45:50.20 & -14:51:11.3 & 25.44 & 23.97 & 23.18 & 23.40 &  & -15.95 &  & -15.14 & 0.15\\
Ha 15 & 19:45:51.05 & -14:51:23.3 & 26.44 & 24.00 & 23.36 & 23.79 &  & -16.39 &  & -15.22 & 0.07\\
Ha 16 & 19:45:12.08 & -14:52:08.6 & 23.52 & 22.04 & 21.77 & 22.30 &  & -15.11 &  & -14.52 & 0.26\\
Ha 17 & 19:45:01.9 & -14:52:55.9 & 24.58 & 23.07 & 23.51 & 24.16 &  & -15.58 &  & -15.29 & 0.51 & faint PN?\\
Ha 18 & 19:45:08.3 & -14:53:22.0 & 23.85 & 22.28 & 22.68 & 23.12 &  & -15.26 &  & -14.92 & 0.46\\
Ha 19 & 19:44:55.91 & -14:55:06.4 & 24.28 & 22.67 & 23.14 & 23.37 &  & -15.44 &  & -15.12 & 0.47\\
Ha 20 & 19:45:04.16 & -14:55:57.4 & $>$ 26.0 & 23.89 & 24.21 & 25.41 &  & -16.20 &  & -15.59 & 0.25\\
Ha 21 & 19:45:14.85 & -14:58:37.2 & 25.44 & 23.70 & 23.94 & 24.54 &  & -15.95 &  & -15.47 & 0.33& faint PN?\\
\end{longtable}
\end{landscape}
}
\section{Appendix}

\begin{figure}[ht] 
\begin{center}
\label{IDcharts}
\includegraphics[width=12cm,height=14cm]{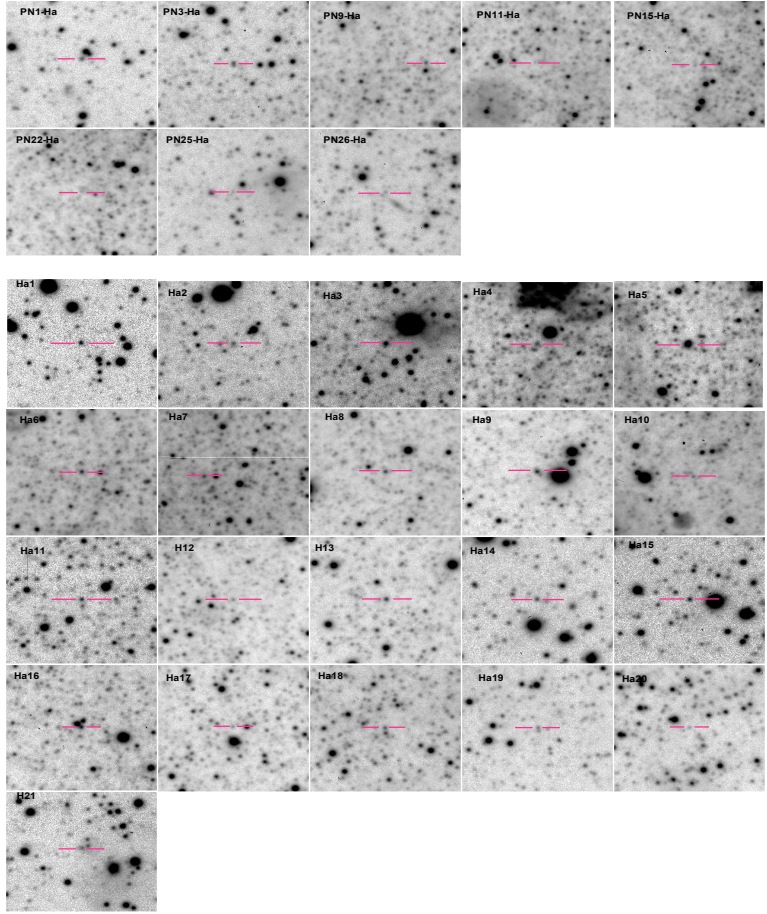}
\caption{ID charts are shown for the newly discovered H$\alpha$ emitting objects. The charts are 30$''$ side and were obtained from the H$\alpha$ image. Each chart is marked with the object number, according to Table 2.}
\end{center}
\end{figure}
\end{document}